\documentclass[a4paper]{jpconf} 
\usepackage{graphicx,amsmath,amssymb,xcolor}

\begin{document}
\title{Searching for phase transitions in neutron stars with modified Gaussian processes}
\author{Debora Mroczek$^1$, M. Coleman Miller$^2$, Jacquelyn Noronha-Hostler$^1$ and Nicolas Yunes$^1$}

\address{
    $^{1}$ {Illinois Center for Advanced Studies of the Universe, Department of Physics, University of Illinois at Urbana-Champaign, Urbana, IL 61801, USA}
} \address{
    $^{2}$ Department of Astronomy, University of Maryland, College Park, 20742, MD, USA
}

\ead{deboram2@illinois.edu}

\begin{abstract}
Gaussian processes provide a promising framework by which to extrapolate the equation of state (EoS) of cold, catalyzed matter beyond $1-2$ times nuclear saturation density.  Here we discuss how to extend Gaussian processes to include nontrivial features in the speed of sound, such as bumps, kinks, and plateaus, which are predicted by nuclear models with exotic degrees of freedom. Using a fully Bayesian analysis incorporating measurements from X-ray sources, gravitational wave observations, and perturbative QCD results, we show that these features are compatible with current constraints and report on how the features affect the EoS posteriors.

\end{abstract}

\section{Introduction}
 Numerous types of electromagnetic and gravitational-wave data sets are now available for neutron stars, and these are being used to infer the composition, thermodynamic, and dynamical properties of the core matter of these stars (see, e.g., Refs. \cite{Pang:2022rzc,Gorda:2022jvk,Han:2021kjx,Raaijmakers:2021uju, Miller:2021qha,Somasundaram:2021clp}). Upgrades to current gravitational wave detectors as well as future instruments and results from the Neutron star Interior Composition Explorer (NICER) mission are expected to improve the data available and, therefore, tighten the constraints on the equation of state (EoS) of cold, catalyzed nuclear matter in neutron stars. 

 An important part of the inference procedure, which is still under active debate, is how to generate minimally biased candidate EoS that incorporate all relevant physics.  Parametric models such as piecewise polytropes or spectral decompositions are commonly used, but these can introduce unwanted correlations across density scales \cite{Legred:2022pyp}. Additionally, sharp features in the speed of sound -- which are present in nuclear models with exotic degrees of freedom (see, e.g. Ref.~\cite{Clevinger:2022xzl}) -- are not well represented in the most common parameterizations, although they play an important role in understanding heavy, ultra-heavy, and twin stars \cite{Tan:2020ics,Tan:2021ahl}. 

 Instead, we start with procedures based on Gaussian processes (GP), which are more model-agnostic and can mitigate these issues at the cost of increased functional complexity \cite{Landry:2018prl}.  It has been shown that GP ensembles capture a wide range of features in the EoS, including phase transitions of varying strength at different densities (see Refs. \cite{Essick:2019ldf,Legred:2021hdx} for detailed discussions). However, investigating specific features in the EoS with GPs can pose computational and sampling challenges. 
 
Here, as a way to produce priors that adequately capture behavior that is predicted by state-of-the-art nuclear models at a low computational cost and without introducing model dependencies, we present a method for generating EoS with sharp features, which we call modified Gaussian processes (mGP). An mGP sample consists of a smooth baseline EoS sampled from a tailored GP that is modified over some range in pressure. The modifications are constructed such that we produce bumps, spikes, oscillations, plateaus, and first-order phase transitions while keeping track of how and where modifications appear. We generate a prior distribution containing samples from both a GP with fixed hyperparameters and a modified GP and construct mass-radius and EoS posteriors using available measurements and theoretical input based on the Bayesian procedure described in Ref.~\cite{Miller:2019nzo}. The different constructions lead to nearly identical mass-radius posteriors, but the EoS posterior is wider in the mGP than in the unmodified GP framework for densities above $\sim$ 1.5 times the nuclear saturation density ($n_{\rm sat} = 0.15 \pm 0.01\textrm{ fm}^{-3}$). In addition, the Bayesian evidence for these non-smooth EoS is indistinguishable from that for smooth EoS.  Thus, current measurements are consistent with sharp and non-trivial features in the EoS, but the data is not yet informative enough to favor or disfavor them.

\section{Generating the Equation of State}

In this section, we give a brief review of GP EoS and describe how we perform our modifications. 

Generally, a Gaussian process yields the joint probability density for a continuous function $f(x)$ at domain points $\vec{x} = \{x_i\} $, which is a tool for approximating $f$ over some domain. This joint probability density is presumed to be a multivariate Gaussian with means $\vec{\mu}$ and covariance matrix $\Sigma$ for the set of domain points $\vec{x}$. That is,
\begin{align}
    f(\vec{x}) \sim \mathcal{N}(\vec{\mu},\Sigma),
\end{align}
where $\mathcal{N}$ indicates a normal distribution. The means and covariance matrix are specified by hyperparameters and depend on further assumptions about $f(x)$.

In this context, a given functional form for the neutron star matter EoS can be sampled from a GP which approximates the relationship between two thermodynamic quantities. We follow Refs.~\cite{Lindblom:2010bb,Landry:2018prl} in introducing the auxiliary variable 
\begin{align}
    \phi \equiv \ln \left(\dfrac{d\varepsilon}{d P} - 1\right),
\end{align}
where $\epsilon$ is the energy density and $P$ is the pressure.  For any $\phi\in (-\infty,\infty)$, the adiabatic speed of sound, $c_s^2 = dP/d\varepsilon$, is stable and causal by construction, since when $\phi \rightarrow +\infty$, $c_s^2 \rightarrow 0$, and when $\phi \rightarrow -\infty$, $c_s^2 \rightarrow 1$, in units where the speed of light $c\equiv 1$. 

We model $\phi$ as a function of $\log_{10} P$ in cgs units at 100 equally spaced points in the range $\log_{10}P (\textrm{erg cm$^{-3}$}) \sim 32 - 38$, using the following trend for the means
\begin{align}\label{eq:eosansatz}
    \mu_i(\log_{10}P_i) = b -2(\log_{10}P_i - 32.7),
\end{align}
where $b = 5.5$ or 3.7 (sampled with equal probability). The first value of $b$ corresponds to the approximate trend found in Ref. \cite{Miller:2021qha} using several tabulated nuclear EoS. This trend largely produces EoS that approach the causal limit in the range of maximal central densities. The second value of $b$ produces EoS that tend to much lower values of $c_s^2$ and are near the conformal limit $c_s^2 \rightarrow 1/3$ at central densities for maximally massive stars.

We assume that the correlation between the function values at different domain points can be represented by a kernel function of the two points, $\Sigma (f_i,f_j) = K(x_i,x_j)$. We also assume that $K(x_i,x_j)$ is a Gaussian which depends only on the distance between the two points, commonly referred to as the squared-exponential kernel,
\begin{align}
    K_{\textrm{se}}(x_i,x_j) = \sigma^2\exp{\left[-\dfrac{(x_i - x_j)^2}{2l^2}\right]},
\end{align}
where $l$ determines the correlation length scale (the limit where all points are independent of each other is $l \rightarrow 0$ ) and $\sigma$ is the strength of the overall correlation ($\sigma \rightarrow 0$ means the variance is negligible). We choose to fix $l=\sigma=1$, following Ref.~\cite{Miller:2021qha}, which is motivated by the spread of the tabulated EoS used to obtain Eq. (\ref{eq:eosansatz}). Our approach differs from that in Ref.~\cite{Miller:2021qha} in that we allow for two values of $b$, which introduces an additional degree of flexibility and ensures that a wider range of $c_s^2$ are represented, despite the other hyperparameters being fixed.  

A modified Gaussian process EoS is produced by taking a sample from the GPs outlined above, which becomes the baseline EoS, removing \textit{part} of the baseline EoS over some range in $P$, and then replacing it with a modification. Modifications are introduced in the form of spikes and plateaus in three different categories: 
\begin{enumerate}
    \item Sharp bump/dip -- a single point in the EoS (selected randomly with uniform probability) gets shifted up/down with respect to the baseline to a random value of $c_s^2$ between 0 and 1 that is at least 10\% above or below the baseline value.
    
    \item Bump/dip + plateau -- In addition to a bump/dip, a plateau is introduced. A plateau spans between 2 and 20 points in pressure (in units of erg cm$^{-3}$, $\log_{10}\Delta P_\textrm{plateau} = 0.12 - 1.2$). The location, height, and length of a plateau are sampled randomly with uniform probability across the allowed ranges in $P$ and $c_s^2$. There is a conditional relationship between the bump/dip and the plateau: if a bump above the baseline value is introduced, the plateau must be below the baseline value at the point where the plateau is introduced. If a dip is introduced, the plateau must be above the baseline value. The 10\% rule of the sharp bump/dip modification still applies. 
    
    \item Double plateau -- In this case two plateaus are introduced, instead of a bump/dip + plateau. The same rules apply.
\end{enumerate}

We produce a total of 900,000 candidate EoS. For any given sample there is a 25\% probability that it will be left unmodified and a 75\% probability that it will be modified in one of the three ways listed above (25\% for each category). At densities below $0.5n_{\rm sat}$ (roughly the crust-core transition density \cite{Hebeler:2013nza}), we use the QHC19 EoS \cite{Baym:2019iky}. We also compute the full range of thermodynamic quantities and remove any mGP samples that contain modifications below the nuclear saturation density, where the properties of equilibrium nuclear matter are better constrained.

\begin{figure*}
   \centering
\begin{tabular}{cc}
\includegraphics[width=.45\linewidth]{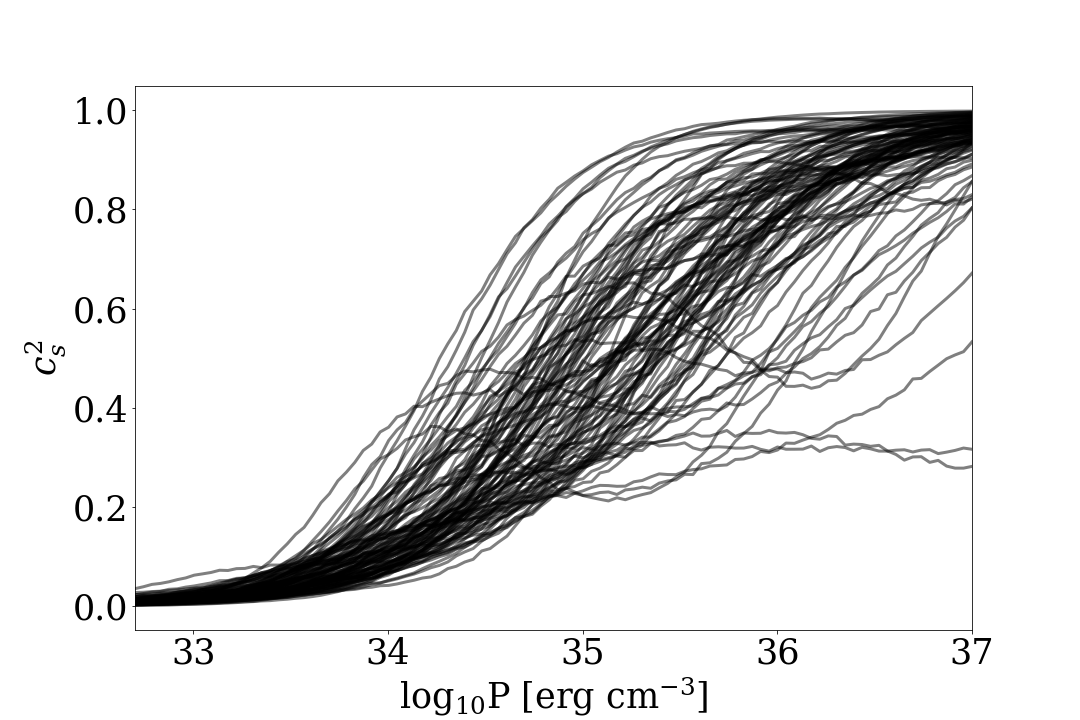} & 
\includegraphics[width=.45\linewidth]{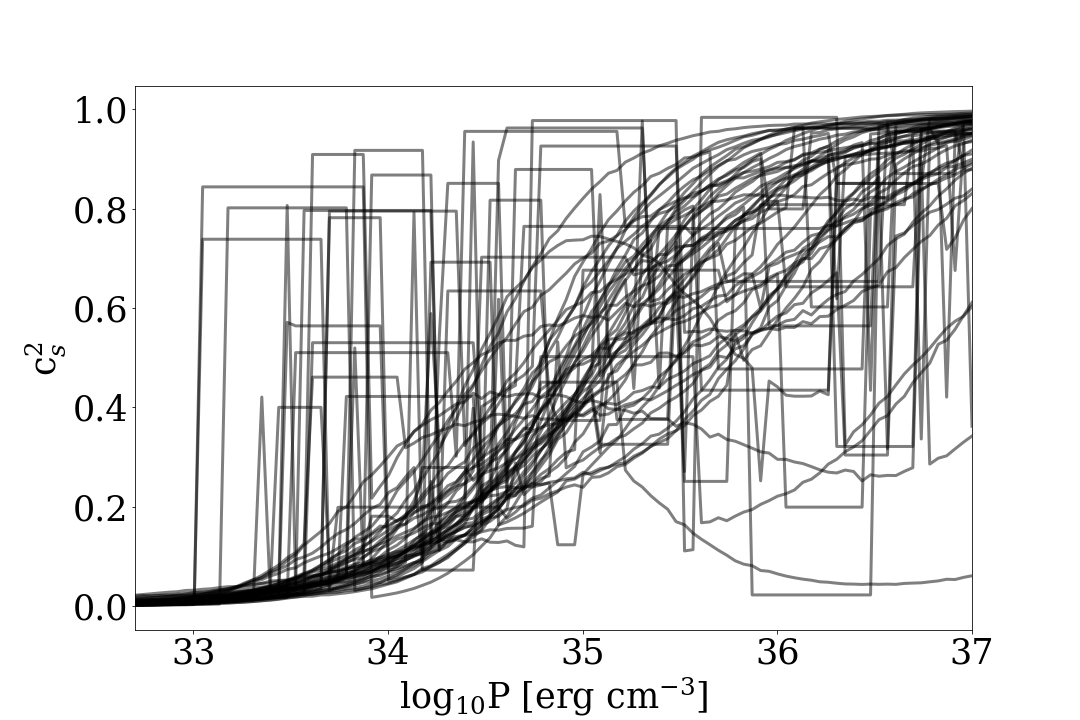} 
\end{tabular}
    \caption{The speed of sound squared in units of $c^2$, as a function of the pressure, for an unmodified (left panel) and modified (right panel) Gaussian process.  We use a squared-exponential kernel with fixed hyperparameters, $\sigma = 1$, $l = 1$, and $\mu(\log_{10} P) = b - 2(\log_{10} P - 32.7)$, where $P$ is the pressure in cgs units and $b = 5.5$ or 3.7 (see text for more details). These samples have not been compared with data, which means that the EoS need not be consistent with observational and theoretical constraints.}
    \label{fig:samples}
\end{figure*}

Figure \ref{fig:samples} illustrates samples from both the Gaussian process and the modified Gaussian process, where the speed of sound is shown as a function of the pressure. Through this small sample of EoS, we see that the modification procedure introduces various interesting features consistent with nuclear models with exotic degrees of freedom, at a low computational cost. We emphasize that traditional GPs can, in principle, approximate these functions and that the unmodified GPs implemented here are tailored to two specific trends and are, thus, not representative of the capabilities of GPs in general. However, the low computational cost and ability to track the location and properties of nontrivial features associated with mGPs make this framework useful for studying exotic degrees of freedom in neutron star cores.

\section{Prior pruning based on observational constraints}

Before proceeding with the Bayesian analysis, we perform a process of informing our prior with conservative cuts motivated by NICER measurements of PSR J0030+0451 \cite{Miller:2019cac} and J0740+6620\cite{Miller:2021qha}, and the detection of GW170817 \cite{de2018tidal}. We refer to this step as \textit{pruning}, as it is used not to extract the EoS, but rather to ensure we have a suitable number of strong candidate EoS in our prior.

The pruning process happens in two steps. We first impose that the maximum mass of a nonrotating star, $M_{\rm max}$, must be at least 1.8 $M_\odot$. For these EoS, we then calculate the equatorial circumferential radius and dimensionless tidal deformability for a 1.4 $M_\odot$ star, $R_{1.4}$ and $\Lambda_{1.4}$, respectively. We require that $10 \leq \Lambda_{1.4} \leq 2000$ and $9.0 \leq R_{1.4} \leq 18.0 $ km. 

\begin{figure*}
\centering 
\includegraphics[width=0.8\linewidth]{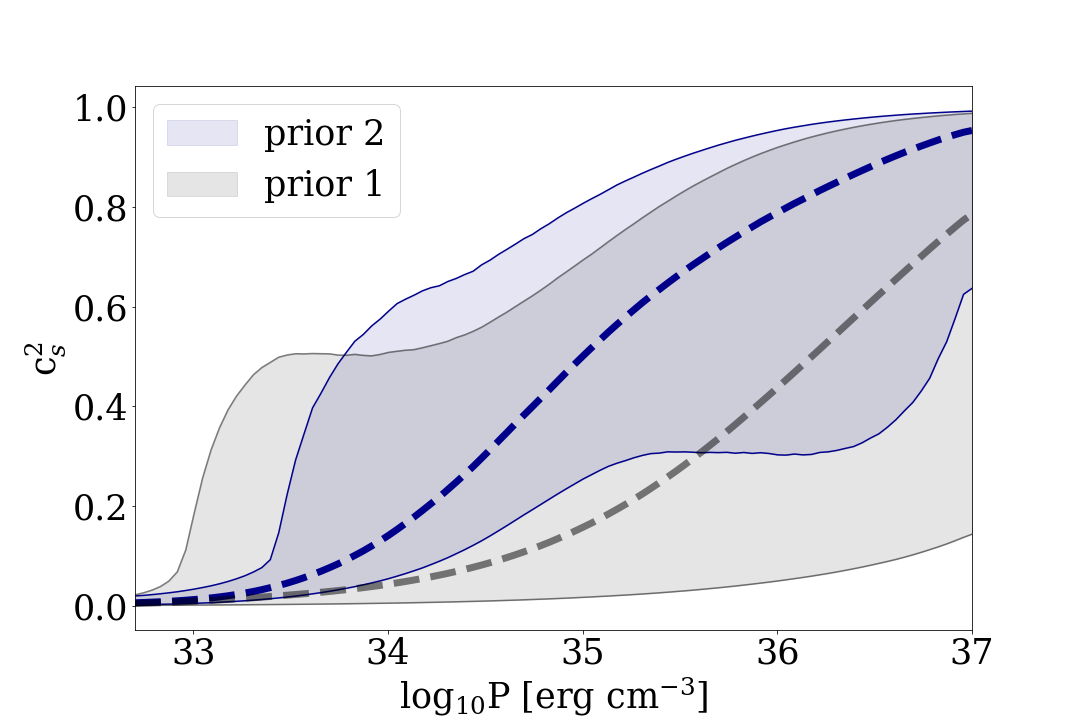}
  \caption{The speed of sound squared in units of $c^2$ as a function of the pressure from the combined distribution of GP and mGP samples. The solid lines delimit the 5\% and 95\% contours and the dashed lines indicate the mean. Prior 1 (gray) corresponds to 900,000 candidate EoS before any pruning is performed. Prior 2 (blue) contains the 104,594 EoS that meet the conservative criteria for the mass and tidal deformability based on GW170817 and PSR J0030+0451 and $M_{\textrm{max}} \geq 1.8 M_\odot$.}
\label{fig:priors}
\end{figure*}

In Figure~\ref{fig:priors}, we show the effect of the prior pruning on the speed of sound of the combined GP and mGP samples. The priors are labeled 1 and 2 and are represented by their means and 90\% contours. Prior 1 is the original sample of 900,000 EoS (25/75 \% split between GP/mGP samples) which does not incorporate assumptions about the physics. Prior 2 shows the 104,594 samples (30/70 \% split between GP/mGP samples) which meet the requirements informed by GW190817 and PSR J0030+0451, the mass cut $M_\textrm{max} \geq 1.8 M_\odot$, and had no modifications below $n_\textrm{sat}$. This final set of EoS is overall stiffer compared to prior 1 but constrained to low values of $c_s^2$ for $\log_{10} P \lesssim 33.5$ because of
the constraints at $1.4 M_\odot$ and theoretical input. We use prior 2 in the Bayesian analysis. 

With Figure~\ref{fig:priors}, we also note that implementing some form of prior pruning is important in nonparametric frameworks for the EoS, especially in the model-agnostic limit. Because these frameworks incorporate minimal physics by construction, creating a reasonable subset of EoS takes significantly more samples. Taking the numbers from our analysis, out of an ensemble of 900,000 EoS, only $\sim10\%$ provide a reasonable match to the observed properties of neutron stars. 

\section{Statistical method and constraints}

 In the Bayesian analysis, we follow the procedure outlined in detail in Ref.~\cite{Miller:2021qha}. As usual, the posterior probability of a given EoS model $k$ is proportional to the product of the prior probability and the likelihood, $p_k \propto q_k \mathcal{L}_k$. The likelihood of a set of observations $(i,j)$ for EoS model $k$ is given by

\begin{align}
    \mathcal{L}_k = \prod_i \left [\prod_{j=1}^{j(i)} \mathcal{L}_k(i,j) \right ]
\end{align}
where $i$ is a type of measurement (e.g. mass, radius) and $j$ is an independent measurement of type $i$ (e.g. two independent measurements of the mass of one object). We follow the procedure developed in Ref.~\cite{Miller:2019nzo} to determine $\mathcal{L}_k$ for each EoS, which uses the full posterior probability distributions of all measurements included. 

The constraints included in the analysis are the masses of three high-mass pulsars \cite{arzoumanian2018nanograv, antoniadis2013massive,cromartie2020relativistic}, the joint mass and tidal deformability posteriors extracted from GW170817 \cite{de2018tidal} and GW190425 \cite{LIGOScientific:2020aai}, and the mass-radius posterior from NICER data on PSR J0030-0451 \cite{Miller:2019cac} and PSR J0740+6620 \cite{Miller:2021qha} (see, respectively, \cite{Riley:2019yda} and \cite{Riley:2021pdl} for independent analyses of these two pulsars from a separate group within the NICER collaboration), with a Gaussian prior for the value of the symmetry energy at nuclear saturation density, $S = 32 \pm 2$ MeV \cite{Tsang:2012se}. We also implement perturbative QCD (pQCD) stability and integral constraints from Ref.~\cite{Komoltsev:2021jzg} for the renormalization parameter range X = [1/2,2], using the pQCD likelihood introduced in Ref.~\cite{Gorda:2022jvk} at the central density for a maximally massive star, $n_B^{max}$ (varies for each EoS), as was done in \cite{Somasundaram:2022ztm}.

\section{Results}

We separate the EoS in the prior between modified and unmodified samples and look at the Bayesian evidence (the probability for a model given the constraints) for each model separately. Since we assume each EoS has an equal prior probability, the Bayesian evidence for a smooth EoS (represented by the GP samples) or an EoS with sharp/non-trivial features (mGP samples) is given by the average likelihood of the EoS in each category,

\begin{equation}\label{eq:evidence}
    P[m(GP)| \textrm{constraints}] = \dfrac{1}{N_{m(GP)}}\sum^{N_{(m)GP}}_{i=0} \mathcal{L}_i,
\end{equation}
where the prior probability of a model is $P[m(GP)] = \frac{1}{N_{m(GP)}}$ and $\mathcal{L}_i$ is the likelihood of EoS $i$. 

The Bayes factor $K$ is the ratio between the evidence for hypotheses -- the bigger the deviation from unity, the more indication that a set of data supports one hypothesis over the other. Comparing the evidence for GP EoS against mGP EoS yields

\begin{equation}\label{eq:bayesfac}
    K = \dfrac{P[(GP)| \textrm{constraints}]}{P[(mGP)| \textrm{constraints}]} = 1.126, 
\end{equation}
from which we conclude that sharp and non-trivial features in the EoS are not ruled out by current constraints. Note also that K=1.126 is a statistically insignificant deviation from unity. That is, each hypothesis is comparably good at explaining the data.   

\begin{figure*}[tb]
   \centering
   \begin{tabular}{c}
        \includegraphics[width=0.8\linewidth]{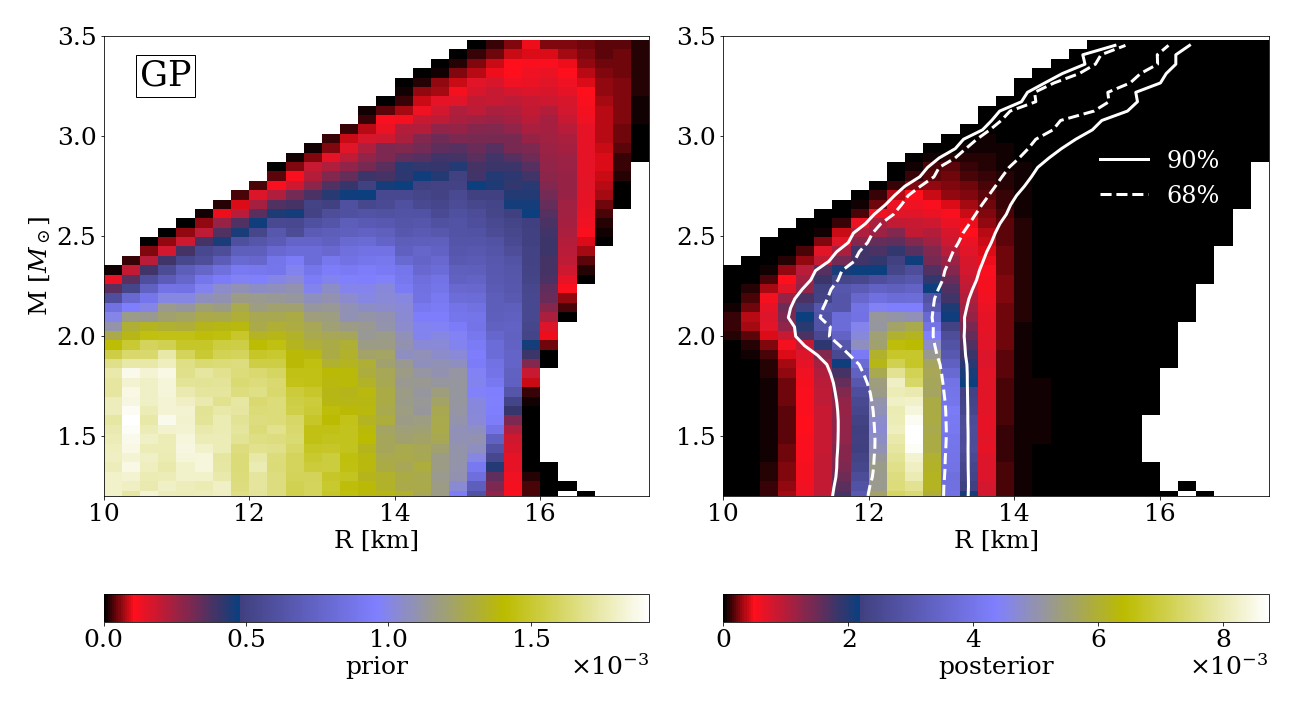}\\
        \includegraphics[width=0.8\linewidth]{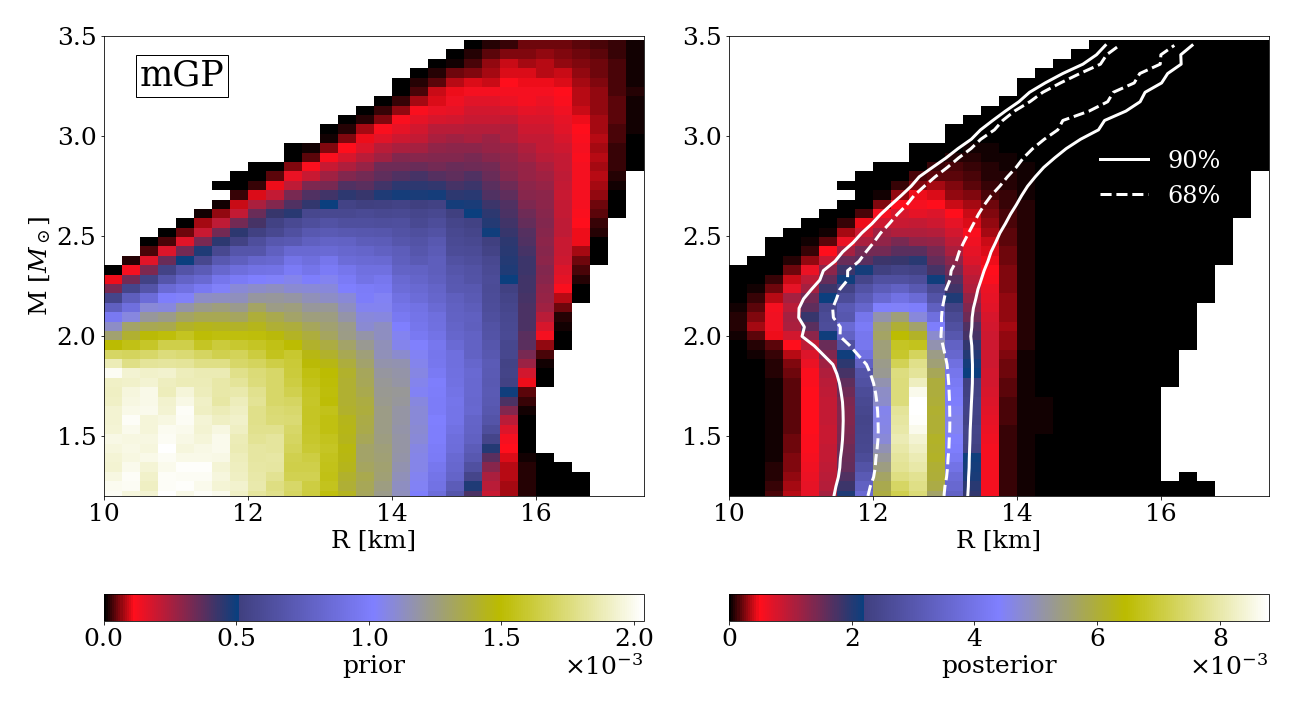}
   \end{tabular}
  \caption{Mass-radius prior (left) and posterior (right) probability distributions for GP (top) and mGP (bottom) EoS. Both the prior and posterior probability distributions are produced by binning the EoS by mass and radius and then normalizing the heights of the bins such that their sum is equal to one. For the posteriors, each EoS is weighted by the corresponding likelihood. Also shown in the posterior plots are the 90\% and 68\% credible regions for the radius at a given mass for $1.1 \leq M \leq 3.5 \ M_{\odot}$.}
  \label{fig:massradius}
\end{figure*} 

In Figure~\ref{fig:massradius}, we show the mass-radius prior and posterior probability distributions for GP and mGP EoS separately and find that they are virtually the same. Rather than approximating a continuous distribution from the samples, we bin the EoS by mass and radius and then normalize such that the sum of the heights of the bins is equal to one. To produce the posterior, we weigh each EoS by its corresponding likelihood before normalizing the bins. We also show the 90\% and 68\% credible regions for the radius at a given mass for $1.1 \leq M \leq 3.5 \ M_{\odot}$. Note that, although we show the credible regions for masses as high as $3.5 \ M_{\odot}$, the posterior probability is negligible for masses $ M \gtrsim 2.7 \ M_{\odot}$. We emphasize that the radius contours shift to the right for large masses because larger maximum masses require stiffer EoS, and thus larger radii. Hence, the shape of the contours is not representative of individual EoS, because the only EoS that contribute in the high-mass region are the ones that reach high masses.

\begin{figure*}
   \centering
   \begin{tabular}{c}
        \includegraphics[width=0.8\linewidth]{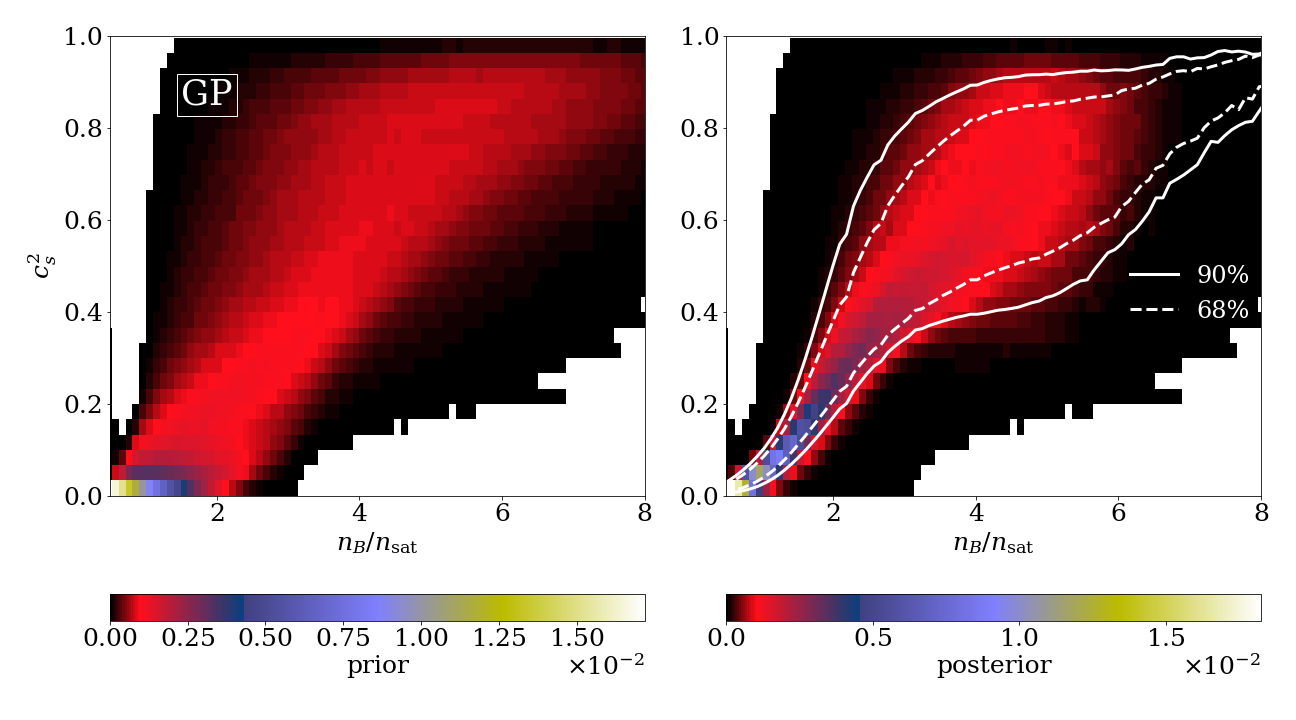}\\
        \includegraphics[width=0.8\linewidth]{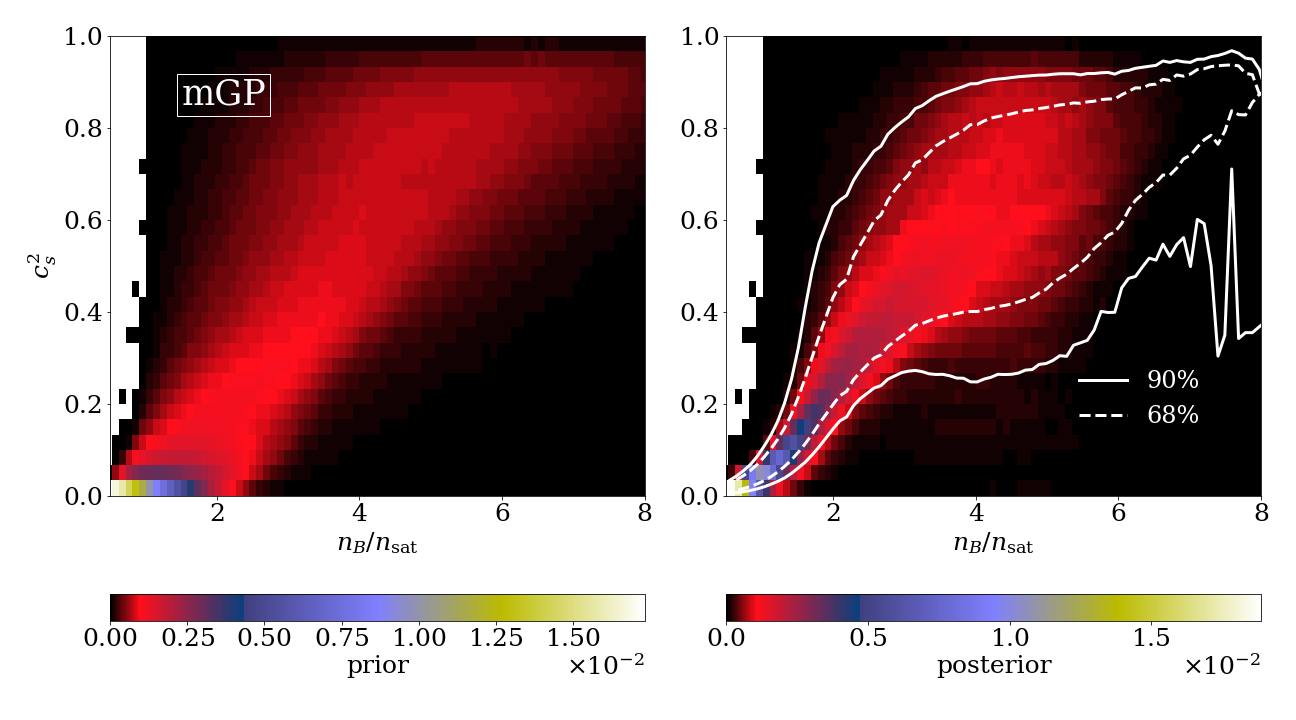}
   \end{tabular}
  \caption{EoS prior (left) and posterior (right) probability distributions for GP (top) and mGP (bottom) EoS. The EoS are represented by the speed of sound squared in units of $c^2$ as a function of baryon number density in units of $n_\textrm{sat}$. The prior and posterior probability distributions are produced by binning the EoS by the speed of sound and number density and then normalizing the heights of the bins such that their sum is equal to one. For the posteriors, each EoS is weighed by the corresponding likelihood. Also shown in the posterior plots are the 90\% and 68\% credible regions for the speed of sound squared at a given density for $0.5 \leq n_B \leq 8.0 \ n_{\rm sat}$. The posterior probability that the central density for a maximally massive star is greater than $\sim6.0 \ n_{\rm sat}$ is negligible in both cases.}
  \label{fig:cs}
\end{figure*}

In Figure~\ref{fig:cs}, we show the EoS posteriors for the square of the speed of sound as a function of baryon number density in units of $n_{\rm sat}$, up to $n_B^\textrm{max}$ for each EoS. The binning and calculation of the prior and posterior probability distributions are done as discussed for Figure \ref{fig:massradius}. Also shown are the 68\% and 90\% credible regions for the speed of sound at a given density up to $n_B = 8.0 \ n_{\rm sat}$, although the posterior probability for $n_B^{max} \gtrsim 6.0 \ n_{\rm sat}$ is negligible. As before, the only EoS that contribute at any given density are the ones that are still stable neutron stars at that particular density. 

Overall, the posteriors are in good agreement. We note that, at the 90\% level, compared to the GP posterior, the mGP posterior is wider. Notably, the mGP posterior allows for slightly stiffer EoS in the regime $1.5 \lesssim n_B \lesssim 3.0\ n_{\rm sat}$, and softer EoS above $3.0\ n_{\rm sat}$. For instance, at twice nuclear saturation density, we extract $c^2_s =0.29^{+0.27}_{-0.11}$ using EoS from the GP and $c^2_s =0.29^{+0.34}_{-0.14}$ using the mGP posterior. At four times nuclear saturation density, the GP EoS range is $c^2_s =0.63^{+0.27}_{-0.23}$, while the mGP EoS allows for $c^2_s =0.59^{+0.31}_{-0.34}$. While the relative differences in the posteriors are dependent on our choices for the two GPs implemented in the unmodified case, the point holds that when the speed of sound is allowed to display sharp features, the posteriors are wider than when a smooth EoS is presumed. We, therefore, argue that such features should be adequately represented in priors for the extraction of the EoS in neutron star regimes. We emphasize that adequate representation requires not only that these features are present in the synthetic EoS, but that additional steps are taken to ensure a sufficient number of EoS with nontrivial features can account for observations (see discussion in Section 3 regarding prior pruning).

\section{Discussion}

We introduced modified Gaussian processes as a new technique for generating synthetic equations of state for the cold, catalyzed nuclear matter in neutron stars. These EoS can be generated at a low computational cost, and contain non-trivial structure consistent with the emergence of exotic degrees of freedom. We prune our samples to ensure that a prior of synthetic EoS contains a large number of strong candidates, given that model-agnostic methods contain minimal physical input by construction. In this analysis, only $\sim10\%$ of the EoS generated in a combined ensemble of modified and unmodified Gaussian process samples were candidates for explaining current mass, radius, and tidal deformability measurements, reinforcing the importance of informing the prior. 

We implement the modified Gaussian process framework in a fully Bayesian analysis to verify that sharp and non-trivial features are consistent with current measurements and theoretical constraints by comparing the evidence for modified Gaussian process EoS against smooth EoS from a standard Gaussian process. Additionally, at the 90\% level, the inclusion of non-trivial structure broadens the allowed range for the speed of sound in the core of neutron stars compared to the assumption that the EoS is smooth. 

The results presented here suggest that current constraints are not enough to rule definitively in favor of or against phase transitions to exotic degrees of freedom in the core of neutron stars. The question of how future measurements can further constrain the nuclear EoS has been explored (see, e.g., Refs. \cite{Landry:2020vaw,Ecker:2022dlg}), but unambiguous signatures of structure in the EoS are still being investigated. Recently, the slope of binary Love relations has been shown to encode structure in the EoS below $3\ n_{\rm sat}$, with twin stars leading to characteristic signatures that can be detected with future instruments~\cite{Tan:2021nat}. However, a more complete picture of the phase structure of QCD for asymmetric matter at zero temperature will require constraints from both astrophysical observations and laboratory measurements, as well as input from effective theories and pQCD (for a detailed discussions see Refs.~\cite{Lovato:2022vgq,Sorensen:2023zkk}).



\ack
D.M is supported by the National Science Foundation Graduate Research Fellowship Program under Grant No. DGE – 1746047, the Illinois Center for Advanced Studies of the Universe Graduate Fellowship, and the University of Illinois Graduate College Distinguished Fellowship. 
J.N.H, D.M., and N.Y.\ were supported in part by the National Science Foundation (NSF) within the framework
of the MUSES collaboration, under grant number OAC2103680. J.N.H. acknowledges financial support from the US-DOE Nuclear Science Grant
No. DESC0020633.  M.C.M. was supported in part by NASA ADAP grant 80NSSC21K0649.
N.Y. was supported in part by NSF Award PHYS-2207650.
The authors also acknowledge support
from the Illinois Campus Cluster, a computing resource that
is operated by the Illinois Campus Cluster Program (ICCP)
in conjunction with the National Center for Supercomputing
Applications (NCSA), which is supported by funds from
the University of Illinois at Urbana-Champaign. 

\section*{References}
\bibliographystyle{unsrt}
\bibliography{ref}

\begin{thebibliography}{10}

\bibitem{Pang:2022rzc}
Peter T.~H. Pang et~al.
\newblock {NMMA: A nuclear-physics and multi-messenger astrophysics framework
  to analyze binary neutron star mergers}.
\newblock 5 2022.

\bibitem{Gorda:2022jvk}
Tyler Gorda, Oleg Komoltsev, and Aleksi Kurkela.
\newblock {Ab-initio QCD calculations impact the inference of the
  neutron-star-matter equation of state}.
\newblock 4 2022.

\bibitem{Han:2021kjx}
Ming-Zhe Han, Jin-Liang Jiang, Shao-Peng Tang, and Yi-Zhong Fan.
\newblock {Bayesian Nonparametric Inference of the Neutron Star Equation of
  State via a Neural Network}.
\newblock {\em Astrophys. J.}, 919(1):11, 2021.

\bibitem{Raaijmakers:2021uju}
G.~Raaijmakers, S.~K. Greif, K.~Hebeler, T.~Hinderer, S.~Nissanke, A.~Schwenk,
  T.~E. Riley, A.~L. Watts, J.~M. Lattimer, and W.~C.~G. Ho.
\newblock {Constraints on the Dense Matter Equation of State and Neutron Star
  Properties from NICER\textquoteright{}s Mass\textendash{}Radius Estimate of
  PSR J0740+6620 and Multimessenger Observations}.
\newblock {\em Astrophys. J. Lett.}, 918(2):L29, 2021.

\bibitem{Miller:2021qha}
M.~C. Miller et~al.
\newblock {The Radius of PSR J0740+6620 from NICER and XMM-Newton Data}.
\newblock {\em Astrophys. J. Lett.}, 918(2):L28, 2021.

\bibitem{Somasundaram:2021clp}
Rahul Somasundaram, Ingo Tews, and J\'er\^ome Margueron.
\newblock {Investigating signatures of phase transitions in neutron-star
  cores}.
\newblock {\em Phys. Rev. C}, 107(2):025801, 2023.

\bibitem{Legred:2022pyp}
Isaac Legred, Katerina Chatziioannou, Reed Essick, and Philippe Landry.
\newblock {Implicit correlations within phenomenological parametric models of
  the neutron star equation of state}.
\newblock {\em Phys. Rev. D}, 105(4):043016, 2022.

\bibitem{Clevinger:2022xzl}
A.~Clevinger, J.~Corkish, K.~Aryal, and V.~Dexheimer.
\newblock {Hybrid equations of state for neutron stars with hyperons and
  deltas}.
\newblock {\em Eur. Phys. J. A}, 58(5):96, 2022.

\bibitem{Tan:2020ics}
Hung Tan, Jacquelyn Noronha-Hostler, and Nico Yunes.
\newblock {Neutron Star Equation of State in light of GW190814}.
\newblock {\em Phys. Rev. Lett.}, 125(26):261104, 2020.

\bibitem{Tan:2021ahl}
Hung Tan, Travis Dore, Veronica Dexheimer, Jacquelyn Noronha-Hostler, and
  Nicol\'as Yunes.
\newblock {Extreme matter meets extreme gravity: Ultraheavy neutron stars with
  phase transitions}.
\newblock {\em Phys. Rev. D}, 105(2):023018, 2022.

\bibitem{Landry:2018prl}
Philippe Landry and Reed Essick.
\newblock {Nonparametric inference of the neutron star equation of state from
  gravitational wave observations}.
\newblock {\em Phys. Rev. D}, 99(8):084049, 2019.

\bibitem{Essick:2019ldf}
Reed Essick, Philippe Landry, and Daniel~E. Holz.
\newblock {Nonparametric Inference of Neutron Star Composition, Equation of
  State, and Maximum Mass with GW170817}.
\newblock {\em Phys. Rev. D}, 101(6):063007, 2020.

\bibitem{Legred:2021hdx}
Isaac Legred, Katerina Chatziioannou, Reed Essick, Sophia Han, and Philippe
  Landry.
\newblock {Impact of the PSR J0740+6620 radius constraint on the properties of
  high-density matter}.
\newblock {\em Phys. Rev. D}, 104(6):063003, 2021.

\bibitem{Miller:2019nzo}
M.~Coleman Miller, Cecilia Chirenti, and Frederick~K. Lamb.
\newblock {Constraining the equation of state of high-density cold matter using
  nuclear and astronomical measurements}.
\newblock 4 2019.

\bibitem{Lindblom:2010bb}
Lee Lindblom.
\newblock {Spectral Representations of Neutron-Star Equations of State}.
\newblock {\em Phys. Rev. D}, 82:103011, 2010.

\bibitem{Hebeler:2013nza}
K.~Hebeler, J.~M. Lattimer, C.~J. Pethick, and A.~Schwenk.
\newblock {Equation of state and neutron star properties constrained by nuclear
  physics and observation}.
\newblock {\em Astrophys. J.}, 773:11, 2013.

\bibitem{Baym:2019iky}
Gordon Baym, Shun Furusawa, Tetsuo Hatsuda, Toru Kojo, and Hajime Togashi.
\newblock {New Neutron Star Equation of State with Quark-Hadron Crossover}.
\newblock {\em Astrophys. J.}, 885:42, 2019.

\bibitem{Miller:2019cac}
M.~C. Miller et~al.
\newblock {PSR J0030+0451 Mass and Radius from $NICER$ Data and Implications
  for the Properties of Neutron Star Matter}.
\newblock {\em Astrophys. J. Lett.}, 887(1):L24, 2019.

\bibitem{de2018tidal}
Soumi De, Daniel Finstad, James~M Lattimer, Duncan~A Brown, Edo Berger, and
  Christopher~M Biwer.
\newblock Tidal deformabilities and radii of neutron stars from the observation
  of gw170817.
\newblock {\em Physical review letters}, 121(9):091102, 2018.

\bibitem{arzoumanian2018nanograv}
Zaven Arzoumanian, PT~Baker, Adam Brazier, Sarah Burke-Spolaor, SJ~Chamberlin,
  Shami Chatterjee, Brian Christy, James~M Cordes, Neil~J Cornish, Fronefield
  Crawford, et~al.
\newblock The nanograv 11 year data set: pulsar-timing constraints on the
  stochastic gravitational-wave background.
\newblock {\em The Astrophysical Journal}, 859(1):47, 2018.

\bibitem{antoniadis2013massive}
John Antoniadis, Paulo~CC Freire, Norbert Wex, Thomas~M Tauris, Ryan~S Lynch,
  Marten~H Van~Kerkwijk, Michael Kramer, Cees Bassa, Vik~S Dhillon, Thomas
  Driebe, et~al.
\newblock A massive pulsar in a compact relativistic binary.
\newblock {\em Science}, 340(6131):1233232, 2013.

\bibitem{cromartie2020relativistic}
H~Thankful Cromartie, Emmanuel Fonseca, Scott~M Ransom, Paul~B Demorest, Zaven
  Arzoumanian, Harsha Blumer, Paul~R Brook, Megan~E DeCesar, Timothy Dolch,
  Justin~A Ellis, et~al.
\newblock Relativistic shapiro delay measurements of an extremely massive
  millisecond pulsar.
\newblock {\em Nature Astronomy}, 4(1):72--76, 2020.

\bibitem{LIGOScientific:2020aai}
B.~P. Abbott et~al.
\newblock {GW190425: Observation of a Compact Binary Coalescence with Total
  Mass $\sim 3.4 M_{\odot}$}.
\newblock {\em Astrophys. J. Lett.}, 892(1):L3, 2020.

\bibitem{Riley:2019yda}
Thomas~E. Riley et~al.
\newblock {A $NICER$ View of PSR J0030+0451: Millisecond Pulsar Parameter
  Estimation}.
\newblock {\em Astrophys. J. Lett.}, 887(1):L21, 2019.

\bibitem{Riley:2021pdl}
Thomas~E. Riley et~al.
\newblock {A NICER View of the Massive Pulsar PSR J0740+6620 Informed by Radio
  Timing and XMM-Newton Spectroscopy}.
\newblock {\em Astrophys. J. Lett.}, 918(2):L27, 2021.

\bibitem{Tsang:2012se}
M.~B. Tsang et~al.
\newblock {Constraints on the symmetry energy and neutron skins from
  experiments and theory}.
\newblock {\em Phys. Rev. C}, 86:015803, 2012.

\bibitem{Komoltsev:2021jzg}
Oleg Komoltsev and Aleksi Kurkela.
\newblock {How Perturbative QCD Constrains the Equation of State at
  Neutron-Star Densities}.
\newblock {\em Phys. Rev. Lett.}, 128(20):202701, 2022.

\bibitem{Somasundaram:2022ztm}
Rahul Somasundaram, Ingo Tews, and J\'er\^ome Margueron.
\newblock {Perturbative QCD and the Neutron Star Equation of State}.
\newblock 4 2022.

\bibitem{Landry:2020vaw}
Philippe Landry, Reed Essick, and Katerina Chatziioannou.
\newblock {Nonparametric constraints on neutron star matter with existing and
  upcoming gravitational wave and pulsar observations}.
\newblock {\em Phys. Rev. D}, 101(12):123007, 2020.

\bibitem{Ecker:2022dlg}
Christian Ecker and Luciano Rezzolla.
\newblock {Impact of large-mass constraints on the properties of neutron
  stars}.
\newblock {\em Mon. Not. Roy. Astron. Soc.}, 519(2):2615--2622, 2022.

\bibitem{Tan:2021nat}
Hung Tan, Veronica Dexheimer, Jacquelyn Noronha-Hostler, and Nicolas Yunes.
\newblock {Finding Structure in the Speed of Sound of Supranuclear Matter from
  Binary Love Relations}.
\newblock {\em Phys. Rev. Lett.}, 128(16):161101, 2022.

\bibitem{Lovato:2022vgq}
Alessandro Lovato et~al.
\newblock {Long Range Plan: Dense matter theory for heavy-ion collisions and
  neutron stars}.
\newblock 11 2022.

\bibitem{Sorensen:2023zkk}
Agnieszka Sorensen et~al.
\newblock {Dense Nuclear Matter Equation of State from Heavy-Ion Collisions}.
\newblock 1 2023.

\end{thebibliography}
 


\end{document}